\documentclass[twocolumn]{aastex63}

\usepackage{graphicx}
\usepackage[T1]{fontenc}
\usepackage{aecompl}
\usepackage[]{amsmath}
\usepackage{color}
\usepackage{xspace}
\usepackage{soul}

\usepackage{lineno}
\nolinenumbers

\newcommand{\msun}{\mathrm{M}_\odot}
\graphicspath{{figures/}}

\def\lsim{ \lower .75ex \hbox{$\sim$} \llap{\raise .27ex \hbox{$<$}} }

\submitjournal{ApJ}

\shorttitle{}
\shortauthors{Tan et al.}
\graphicspath{{./}{figures/}}
\begin{document}

\title{The mass and redshift dependence of halo star clustering}
\correspondingauthor{Wenting Wang}
\email{wenting.wang@sjtu.edu.cn}

\author{Zhenlin Tan}
\affiliation{Department of Astronomy, Shanghai Jiao Tong University, Shanghai 200240, China}
\affiliation{Shanghai Key Laboratory for Particle Physics and Cosmology, Shanghai 200240, China}
\author{Wenting Wang}
\affiliation{Department of Astronomy, Shanghai Jiao Tong University, Shanghai 200240, China}
\affiliation{Shanghai Key Laboratory for Particle Physics and Cosmology, Shanghai 200240, China}
\author{Jiaxin He}
\affiliation{Department of Astronomy, Shanghai Jiao Tong University, Shanghai 200240, China}
\affiliation{Shanghai Key Laboratory for Particle Physics and Cosmology, Shanghai 200240, China}
\author{Yike Zhang}
\affiliation{Department of Astronomy, Shanghai Jiao Tong University, Shanghai 200240, China}
\affiliation{Shanghai Key Laboratory for Particle Physics and Cosmology, Shanghai 200240, China}
\author{Vicente Rodriguez-Gomez}
\affiliation{Instituto de Radioastronom\'ia y Astrof\'isica, Universidad Nacional Aut\'onoma de M\'exico, Apdo. Postal 72-3, 58089 Morelia, Mexico}
\author{Jiaxin Han}
\affiliation{Department of Astronomy, Shanghai Jiao Tong University, Shanghai 200240, China}
\affiliation{Shanghai Key Laboratory for Particle Physics and Cosmology, Shanghai 200240, China}
\author{Zhaozhou Li}
\affiliation{Department of Astronomy, Shanghai Jiao Tong University, Shanghai 200240, China}
\affiliation{Shanghai Key Laboratory for Particle Physics and Cosmology, Shanghai 200240, China}
\author{Xiaohu Yang}
\affiliation{Department of Astronomy, Shanghai Jiao Tong University, Shanghai 200240, China}
\affiliation{Shanghai Key Laboratory for Particle Physics and Cosmology, Shanghai 200240, China}

%\collaboration{1}{(AAS Journals Data Scientists collaboration)}



\begin{abstract}

We adopt the two point correlation function (2PCF) as a statistical tool to quantify the spatial clustering of halo stars, for galaxy systems spanning a wide range in host halo virial mass ($11.25<\log_{10}M_{200c}/\msun<15$) and redshifts ($0<z<1.5$) from the IllustrisTNG  simulations. Consistent with a previous study \cite[][Paper I]{2024ApJ...961..223Z}, we identify clear correlations between the strength of the 2PCF signals and galaxy formation redshifts, but over a much wider mass range. We find that such correlations are slightly stronger at higher redshifts, and get weakened with the increase of host halo mass. We demonstrate that the spatial clustering of halo stars is affected by two factors: 1) the clustering gets gradually weakened as time passes (phase mixing); 2) newly accreted stars at more recent times would increase the clustering. For more massive galaxy systems, they assemble late and the newly accreted stars would increase the clustering. The late assembly of massive systems may also help to explain the weaker correlations between the 2PCF signals and the galaxy formation redshifts in massive halos, as their 2PCFs are affected more by recently accreted stars, while formation redshift characterizes mass accretion on a much longer timescale. We find that the orbits of satellite galaxies in more massive halos maintain larger radial anisotropy, reflecting the more active accretion state of their hosts while also contributing to their stronger mass loss rates.

\end{abstract}



\section{Introduction}
\label{sec:intro}

In the standard hierarchical structure formation theory of our Universe, galaxies are formed within dark matter halos. Smaller halos and galaxies form first, and they merge to form larger galaxies and halos. Once a small halo with stars is accreted by a larger dark matter halo, it gradually loses its bound dark matter and stars due to tidal effects. Stripped stars form coherent stellar streams at first, which then gradually become more diffuse in phase space, eventually contributing to the growth of the central galaxy. In this scenario, the stellar halos of galaxies are formed by the accretion and the disruption of smaller galaxies \citep[e.g.][]{2001ApJ...548...33B}. The existence of tidal streams in both real and velocity space has long been observed around our Milky Way (hereafter MW) \citep[e.g.][]{2000ApJ...540..825Y,2001ApJ...547L.133I,2002ApJ...569..245N,2003ApJ...588..824Y,2003MNRAS.340L..21I,2006ApJ...642L.137B,2007ApJ...668..221N}, in the nearby Andromeda galaxy \citep[e.g.][]{2001Natur.412...49I,2004MNRAS.351..117I,2018NatAs...2..737D} and in other more distant galaxies \citep[e.g.][]{2019ApJ...883...19A,2020AJ....159..103K}. 

It has been recognized that dark matter halos and galaxies have varied accretion histories, and according to the above hierarchical structure formation scenario, the accretion events are imprinted in the phase-space distribution of stars in the stellar halos (halo stars) at today \citep[e.g.][]{1993ARA&A..31..575M,2016ApJ...821....5D,2018MNRAS.474.5300D,2018NatAs...2..737D}. In addition to the most prominent and coherent tidal streams which still survive at today, evidences about accretions happened in the past can be identified by looking for substructures in real, velocity and action space, sometimes using metallicity as a check. For example, by investigating the radial and tangential velocity distribution of stars, our MW was reported to have merged with another massive galaxy at about 8-11~Gyr ago \citep[][]{2018Natur.563...85H,2018MNRAS.478..611B}. In fact, dozens of methods have been applied to look for smaller substructures through, e.g., matched filters, entropy study, clustering algorithms, tomography and machine learning \citep[e.g.][]{1999Natur.402...53H,2002AJ....124..349R,2006ApJ...639L..17G,2009ApJ...698..567S,2011ApJ...738...79X,2013ApJ...769L..23G,2014ApJ...790L..10G,2016MNRAS.463.1759B,2018ApJ...863...26Y,2018MNRAS.478.5449M,2020ApJ...891...39Y,2020ApJ...904...61Z,2022ApJ...930...23N,2024ApJ...965...10T}. Beyond our MW, orbital superposition models fitting to the observed surface brightness, stellar kinematics, age, and metallicity maps based on Integral Field Unit (IFU) observations of galaxies, can reveal massive merger events happened in the past \citep[e.g.][]{2022A&A...660A..20Z,2022A&A...664A.115Z}.

The aforementioned approaches usually identify individual accretion events, but they do not provide quantitative estimates of the strength of clustering for halo stars. To quantitatively describe the strength of clustering of halo stars, the two point correlation function (hereafter 2PCF) is a possible choice, which has been a powerful statistical tool used to quantify the clustering of galaxies from kiloparsec to megaparsec and even a hundred of megaparsec scales \citep[e.g.][]{2005ApJ...630....1Z,2005MNRAS.362..711Y,2005MNRAS.357..608Y,2009MNRAS.397.1862P,2011ApJ...734...88W,2012MNRAS.424.1471L,2014MNRAS.441.2398G,2013PhRvD..88j3510Z,2007MNRAS.381.1053P,1998ApJ...494....1J,2000MNRAS.318.1144P,2002ApJ...575..587B,2005ApJ...633..791Z,2021MNRAS.505.2784Z,1998ApJ...494....1J,2005MNRAS.362..505C,2023ApJ...948...99Z}. It has also been applied to study the clustering of stars in stellar clusters on subparsec scales \citep[e.g.][]{2017A&A...599A..14J,2018A&A...620A..27J}, used to enable statistical comparisons between real observation and predictions by modern numerical simulations for the structure of MW stellar halo \citep[e.g.][]{2011MNRAS.417.2206C}, and adopted to study the symmetry-breaking pattern \citep{2012ApJ...750L..41W} in the orthogonal and radial directions of the Galactic disk \citep{2023ApJ...942...41H}. 

In particular, a recent study of \cite{2019MNRAS.484.2556L} calculated the halo star 2PCF signals using both real observed MW halo stars and simulated stars. Comparisons were performed between the signals of real observation and simulations, and the authors picked up simulated MW-like galaxy systems which have closer 2PCF signals as those measured from real observed MW halo stars, and the mass assembly history of these simulated systems were used to represent the mass assembly history of our MW. However, using MW-mass galaxies selected from the TNG50 simulation, we find that although there are clear evidences showing definite correlations between the 2PCF signals of halo stars and the mass assembly histories, the scatter is very large \citep[][hereafter Paper I]{2024ApJ...961..223Z}. We report that the 2PCF signals of halo stars can be used to statistically quantify the clustering of halo stars, but using the 2PCF {\it alone} is difficult to precisely predict the assembly history of our MW. 

Our current paper is a follow-up study of Paper I. We continue adopting the halo star 2PCF signals to quantify the phase-space clustering strength and the maintenance of substructures/stellar streams in halo stars of the IllustrisTNG series of simulations. On the basis of Paper I, in which we only looked at MW-mass galaxies, we now investigate how the correlations between the 2PCF signals and the stellar halo formation redshifts\footnote{The redshift when the galaxy has accreted a certain fraction of the total mass for the galaxy at today or at other reference time.} change in host dark matter halos with different masses and at different redshifts. We will discuss physical interpretations on such dependencies, especially the velocity anisotropy in the orbital distributions of satellite galaxies or subhalos in shaping such a halo mass dependence. 

The layout of this paper is as follows. In Section~\ref{sec:data}, we introduce the IllustrisTNG simulations and our galaxy and halo star selections. Section~\ref{sec:methods} details our methods for calculating halo star 2PCF signals and velocity anisotropies for halo stars and satellites galaxies. Section~\ref{sec:results} presents results on the halo mass and redshift dependencies of the correlation between 2PCF signals and galaxy formation redshifts. We perform detailed discussions in Section~\ref{sec:disc} about possible reasons to explain the halo mass dependence. We conclude in the end (Section~\ref{sec:concl}).

\section{Data}
\label{sec:data}

\subsection{The IllustrisTNG Simulation}

 The IllustrisTNG simulations are a suite of hydrodynamical simulations incorporating sophisticated baryonic processes, carried out with the moving-mesh code \citep[\textsc{arepo};][]{Springel2010} to solve the equations of gravity and magneto-hydrodynamics. They include comprehensive treatments of various galaxy formation and evolution processes, such as metal line cooling, star formation and evolution, chemical enrichment and gas recycling. The TNG suites of simulations adopt the Planck 2015 $\Lambda$CDM cosmological model with $\Omega_\mathrm{m}=0.3089$, $\Omega_\Lambda=0.6911$, $\Omega_\mathrm{b}=0.0486$, $\sigma_8=0.8159$, $n_s=0.9667$, and $h=0.6774$ \citep{Planck2015}. A total of 100 snapshots are saved between redshifts of $z=20$ and $z=0$, with the initial condition set up at $z=127$. For more details about TNG, we refer the readers to \cite{Marinacci2018,Naiman2018,Nelson2018,Pillepich2018,Springel2018,Nelson2019}. 

In our analysis, we use the TNG50-1, TNG100-1 and TNG300-1 simulations. They are named in this way because their box sizes are of approximately 50, 100 and 300~Mpc on a side. Explicitly, TNG50-1 is the simulation with the highest resolution in its suite (compared with TNG50-2 and TNG50-3), and hereafter we refer to it as TNG50. It has a periodic box with 35~Mpc/h on a side that follows the joint evolution of 2,160$^3$ dark matter particles and $\sim$2,160$^3$ gas cells. Each dark matter particle has a mass of $3.1\times10^5 \msun$/h, while the baryonic mass resolution is $5.7\times10^4 \msun$/h. Similarly, TNG100-1 and TNG300-1 are the simulations with the highest resolution in their suites, and hereafter we call them TNG100 and TNG300. TNG100 has a periodic box of 75~Mpc/h on a side that follows the joint evolution of $1,820^3$ dark matter particles and $\sim$1,820$^3$ gas cells. Each dark matter particle has a mass of $5.1\times10^6 \msun$/h, while the baryonic mass resolution is $9.4\times10^5 \msun$/h. TNG300 has a periodic comoving box with 205~Mpc/h on each side that follows the joint evolution of 2,500$^3$ dark matter particles and $\sim$2,500$^3$ gas cells. Each dark matter particle has a mass of $4.0\times10^7 \msun$/h, while the baryonic mass resolution is $7.6\times10^6 \msun$/h.

\subsection{Galaxy and halo star Selections}

Our samples of galaxies are selected from the TNG50, TNG100 and TNG300 simulations. We select only halo central galaxies. Moreover, in order to select systems undergoing major mergers that deviate strongly from steady states, we require the absolute $r$-band magnitudes of the central galaxies in each halo to be at least 0.5 magnitudes brighter than their brightest satellites. For each system, we select only ex-situ formed stars\footnote{Ex-situ stars are formed by accreting smaller galaxies, while in-situ stars are formed by gas cooling.} for our analysis, and we call them halo stars or halo star particles hereafter. We also exclude star particles which are still bound to surviving subhalos or satellites and we use the stellar assembly catalogue provided by the TNG website \citep{2016MNRAS.458.2371R,Rodriguez-Gomez2017} to check whether each star particle is ex-situ or in-situ formed. This catalogue is constructed by tracking the baryonic merger trees and it is used to determine whether a stellar particle was formed outside of the main progenitor branch of a given galaxy. If true, it is considered as an ex-situ star particle. Otherwise, the star particle is tagged as an in-situ star particle.

%%%%%%%%%%%%%%%%%%%%%%%%%%%%%%%%%%%%%%%%%%%%%%%%%%%%%%%%%
\begin{table}[ht]
\caption{The bins of $M_{200c}$ over which we calculate the 2PCF signals, and the number of galaxies in each bin, for TNG50, TNG100 and TNG300. For the mass bins marked with *, they overlap across different TNG simulations, and thus can be used to check consistencies with varying resolutions.}
\begin{center}
\begin{tabular}{lcc}\hline
\hline
\multicolumn{1}{c}{Simulation} & \multicolumn{1}{c}{$M_{200c}$ range [$\log_{10}\msun$]}  & \multicolumn{1}{c}{Number} \\ 
\hline
TNG50 & 11.25-11.75 & 229  \\
  & 11.75-12.25 & 187  \\
  & 12.25-12.75* & 71  \\
  & 12.75-15  & 28  \\
\hline
\hline
TNG100 & 12.25-12.75*  & 600 \\  % 244
   & 12.75-13.25  & 193  \\  % 157
   & 13.25-13.75*  & 52  \\  % 43
   & 13.75-15  & 31  \\  % 25
\hline
\hline
TNG300 & 13.25-13.75*  & 1259  \\  % 905
   & 13.75-14.25  & 459  \\  % 383
   & 14.25-15  & 99  \\  % 83
\hline
\label{tbl:massbins}
\end{tabular}
\end{center}
\end{table}
%%%%%%%%%%%%%%%%%%%%%%%%%%%%%%%%%%%%%%%%%%%%%%%%%%%%%%%%%%

In the end, we have 515 galaxies with $\log_{10}M_{200c}/\msun>11.25$ from TNG50. For TNG100, we select 876 galaxies with $\log_{10}M_{200c}/\msun>12.25$. And for TNG300, 1817 galaxies are selected with $\log_{10}M_{200c}/\msun>13.25$. Here $M_{200c}$ is the total mass enclosed within the virial radius, $R_{200c}$, of the host dark matter halos, and $R_{200c}$ is defined as the radius within which the total matter density is 200 times the critical density of the universe. The lower thresholds in $M_{200c}$ we quote for different resolutions of TNG simulations are determined according to the following reasons. Starting from the average mass of star particles in TNG300, TNG100 and TNG50, we require that a realistic satellite galaxy should have at least 100 star particles, which leads to the lower limit in the stellar mass of satellites ($M_{\ast,\mathrm{sat,limit}}$). Halo stars are stripped from bound satellites, which form the stellar halos of central galaxies. To have realistic stellar halos and 2PCF signal measurements of these halo stars, we require that the stellar mass of central galaxies should be at least one order of magnitude higher than $M_{\ast,\mathrm{sat,limit}}$. We then transfer this lower limit in the stellar mass of central galaxies to the lower limit in $M_{200c}$, according to the corresponding stellar mass to halo mass relation in the simulation, and we take the upper bound in $M_{200c}$ at fixed stellar mass. In addition, in order to have robust measurements of the 2PCF signals, we also require that there should be at least 80\% galaxies that can have more than 300 ex-situ star particles with galactocentric distances between $0.1~R_{200c}$ and $R_{200c}$ in each bin of $M_{200c}$ (see below and Table~\ref{tbl:massbins}), and only use these galaxies to calculate the 2PCF signals. This last requirement further raises the mass threshold in $M_{200c}$.

Throughout our analysis in this paper, we not only calculate the 2PCF for galaxies at redshift $z=0$, but also their progenitors at higher redshifts of $z=0.5$, $z=1.0$ and $z=1.5$. The determination of the lower mass thresholds, as described above, is valid for all snapshots used in this study, and for different redshifts, the samples of galaxies in each $M_{200c}$ bin but at different redshifts are ensured to be the same throughout this paper.

The $M_{200c}$ bins we adopt for TNG300, TNG100 and TNG50 are summarized in the second column of Table~\ref{tbl:massbins}, and the third column provides the number of galaxies in each $M_{200c}$ bin. Note that there are overlaps of the same mass bins across different resolutions of TNG simulations, which are marked by those with the $*$ symbol in Table~\ref{tbl:massbins}. These bins can be used for consistency checks among simulations with different resolutions.

\section{Methodology}
\label{sec:methods}

\subsection{Two Point Correlation Function Calculation}

With a continuous density field, the 2PCF, $\xi$, is defined as 

\begin{equation}
\xi(\Delta x)=\langle \delta(x)\delta(x+\Delta x) \rangle,
\label{eqn:2pcfdef1}
\end{equation}

where the average goes over the spatial coordinates, $x$, and $\delta=\frac{\rho}{\langle \rho \rangle}-1$ is the density contrast, which describes the excess in the local density, $\rho$, with respect to the mean density, $\langle \rho \rangle$. 

For a discrete density field, such as the spatial distribution of discrete galaxies or stars, the 2PCF can be equivalently written in the following form

\begin{equation}
\mathrm{d} n_{12}(\Delta x)=\bar{n}_1\bar{n}_2[1+\xi(\Delta x)]\mathrm{d} V_1 \mathrm{d} V_2.
\label{eqn:2pcfdef2}
\end{equation}

$\xi(\Delta x)$ describes the average excess probability, with respect to a uniform distribution, of finding a population 2 object (denoted by lower index 2) with a separation of $\Delta x$ around a population 1 object (denoted by lower index 1). $\mathrm{d} V_1$ and $\mathrm{d} V_2$ in Equation~\ref{eqn:2pcfdef2} above are the local volume element of population 1 and 2 objects. $\bar{n}_1$ and $\bar{n}_2$ are the average spatial number densities of the two populations. $\mathrm{d} n_{12}$ on the left hand side of Equation~\ref{eqn:2pcfdef2} is the pair count of population 1 and 2 objects with spatial separation of $\Delta x$. When population 1 ($\mathrm{D_1}$) and population 2 ($\mathrm{D_1}$) are identical, the 2PCF simply reduces to auto correlation ($\mathrm{D_1}=\mathrm{D_2}=\mathrm{D}$), and throughout this paper, we calculate the auto correlations of ex-situ halo stars. 

In practice, the 2PCF is neither calculated from Equation~\ref{eqn:2pcfdef1} nor Equation~\ref{eqn:2pcfdef2}. Instead, the 2PCF is estimated from the pair counts of data points (in our case the halo star particles) and random points, and the random points are generated following the spatial distributions of data points. In our analysis, we adopt the commonly adopted Landy-Szalay estimator \citep{1993ApJ...412...64L} of the following form to calculate the auto correlations

\begin{equation}
\xi(\Delta x)=\frac{\mathrm{DD}(\Delta x)-2\mathrm{DR}(\Delta x)+\mathrm{RR}(\Delta x)}{\mathrm{RR}(\Delta x)}. 
\label{eqn:estimator}
\end{equation}

Here D stands for the sample of halo star particles from simulations, and R refers to the random sample. DD, DR and RR are the total number of all possible pair counts between halo star pairs, between halo star and random point and between random pairs with a 3-dimensional spatial separation of $\Delta x$. Throughout this paper, the symbol $r$ is adopted to denote the galactocentric radii, and instead we use $\Delta x$ to denote the spatial separations between the pair counts, $\Delta x=\sqrt{(x_1-x_2)^2+(y_1-y_2)^2+(z_1-z_2)^2}$. We choose a fixed range of pair separation, $\Delta x=2.5-5$~kpc for our analysis in this paper. Note in Paper I, we have tried three different pair separations of $\Delta x=0.1-1$~kpc, $\Delta x=1-2.5$~kpc and $\Delta x=2.5-5$~kpc. We do not see significant dependence of our conclusions on the choice of $\Delta x$ in paper I, and thus in this current paper we fix the pair counting separation to $\Delta x=2.5-5$~kpc. In addition, we have tested how the halo star 2PCF signals change with the increase in $\Delta x$, and have seen steady decrease in the 2PCF signal with the increase of $\Delta x$. The signal becomes close to zero when $\Delta x$ is greater than 30-40~kpc.

We first calculate the 2PCF signals using halo stars for each galaxy and in bins of different galactocentric distances. Stars are binned radially first and then the 2PCF signals are calculated separately in each bin. The edge effect of each radial bin can be naturally accounted for, as long as the star particles and random points have exactly the same spatial coverage or bin boundary. As having been pointed out in Paper I, the halo star 2PCFs of individual galaxies have large scatters, and in this paper we will report the overall signal for a group of galaxies binned according to their formations times, host halo masses and redshifts. {\it For galaxies in the same bin, we first calculate the 2PCF signal using halo stars and the corresponding random points for each galaxy, and we report the median of the 2PCF taken over different galaxies in this bin. This is the convention adopted for all results in this paper. } 

In our analysis, the random sample does not have uniform spatial distribution. Instead, they are generated to follow the same radial distribution as the halo stars, but are free of substructures. This is achieved by randomly shuffling the position angles ($\theta$ and $\phi$)\footnote{The position angles are defined through the original right handed Cartesian $X$, $Y$ and $Z$ coordinates in the simulation, with $\theta$ defined as the angle starting from the $X$-axis and counted counter clockwise from 0 to 360 degrees in the $XY$ plane and $\phi$ defined as the angle of $\pm$90 degrees from the $XY$ plane.} of different halo star particles, i.e., we randomly assign the position angle of another particle to the current particle. Note in Paper I, we have tried another method of first fitting a double power law function to the radial number density profiles of halo stars in each galaxy. Based on the best-fit smooth double power law function, we generate random points which follow this radial distribution, but with random angular distributions. This is in fact the approach adopted by \cite{2019MNRAS.484.2556L}, and we have found in Paper I that it leads to fully consistent measurements as our current way of randomly shuffling position angles. Thus in this paper we choose to use the shuffling method, which is easier to implement.

With the random sample generated in this way, our measured 2PCFs in this paper represent the strength of the clustering or clumpiness in halo stars at different galactocentric radii at the fixed pair separation of $\Delta x=2.5-5$~kpc. Note we exclude star particles bound to surviving subhalos or satellites in our analysis, and thus the clumpiness in our sample of halo stars reflects the existence of stellar streams\footnote{Our random samples generated follow spherical symmetry, while the spatial distribution of real halo stars can be triaxial. Thus our 2PCF signal may also contain clustering information due to the large scale triaxial distribution of halo stars, with respect to the spherically distributed random sample, but we expect this would mainly happen at much larger pair separations than our choice of $\Delta x = 2.5-5$~kpc.}. The streams approximately maintain their phase-space clusterings after getting tidally stripped from their progenitor satellite galaxies \citep[e.g.][]{2012MNRAS.420.2700K,2014ApJ...795...95B,2023MNRAS.524.2124P}. For recently stripped stellar streams, which are dynamically cold and more compact, their 2PCF signals will be significantly greater than unity. Because the phase-space clustering of stellar streams becomes weaker over time (phase mixing), the strength in the 2PCF signals correlates with the time/redshift when the merger event happens, as demonstrated in Paper I. 

We adopt the open source code, \textsc{corrfunc} \citep{10.1007/978-981-13-7729-7_1,2020MNRAS.491.3022S}, to calculate the 2PCF signals. \textsc{corrfunc} is a set of fast and high-performance routines to measure clustering statistics. It supports parallel calculations and thus enables efficient counting of the above pairs ($DD$, $DR$ and $RR$) in Equation~\ref{eqn:estimator}.

\subsection{The Velocity Anisotropy Calculation}
\label{sec:beta}

The velocity anisotropy of a population of objects is defined as
\begin{equation}
\beta=1-\frac{\sigma^2_\theta+\sigma^2_\phi}{2\sigma^2_r}
=1-\frac{\langle {v^2_\theta} \rangle- \langle v_\theta \rangle^2+\langle {v^2_\phi}\rangle-\langle v_\phi \rangle^2}{2(\langle {v^2_r} \rangle-\langle v_r \rangle^2)}, 
\label{eqn:beta}
\end{equation}
where $\sigma_r$ is the radial velocity dispersion, while $\sigma_\theta$ and $\sigma_\phi$ are velocity dispersions of the two tangential velocity components. $v_r$ is the radial velocity. $v_\theta$ and $v_\phi$ are the two components of the tangential velocity. $\beta$ can go from minus infinity to 1. When radial orbits dominate the population, $\beta$ is close to 1, and when tangential orbits dominate, $\beta$ is negative. 

We first calculate $\beta$ for each individual galaxy system, and similar to how we report the 2PCFs of galaxies in the same bin, we report the median as well. In this study, we will report the velocity anisotropy for satellites/subhalos in galaxy systems with different host halo masses from TNG50, TNG100 and TNG300, to help interpreting the halo mass dependence of the correlations between 2PCF signals and the galaxy formation redshifts. 

\subsection{The Spearman Correlation Coefficient}
The Spearman Correlation Coefficient, denoted as $\rho$ in this paper, measures the monotonic relationship between two datasets. It quantifies the degree to which the ranking of one variable corresponds to the ranking of another variable, and is defined as
\begin{equation}
\rho = 1 - \frac{6 \sum d^2}{n(n^2 - 1)},
\label{eqn:rho}
\end{equation}
where $d=\mathrm{Rank}(x_{i})-\mathrm{Rank}(y_{i})$. Here $x_{i}$ and $y_{i}$ are the $\mathrm{i}^\mathrm{th}$ values in the corresponding datasets and Rank() denotes the index of the $\mathrm{i}^\mathrm{th}$ value in sorted datasets.
$\rho$ varies between $-$1 and 1, with 0 indicating no correlation between the two datasets. $\rho=-$1 or 1 implies an exact monotonic relationship. Positive correlations indicate that $y$ tends to increase as $x$ increases. Negative correlations indicate that $y$ tends to decrease as $x$ increases.

We calculate an overall coefficient between the 2PCF signals at different radii and galaxy formation redshifts for different galaxies within each mass bin. We use this quantity to demonstrate the relationship between 2PCF signals and galaxy formation time.

\section{Results}
\label{sec:results}

\begin{figure*}[ht]
\centering
\includegraphics[width=0.86\linewidth]{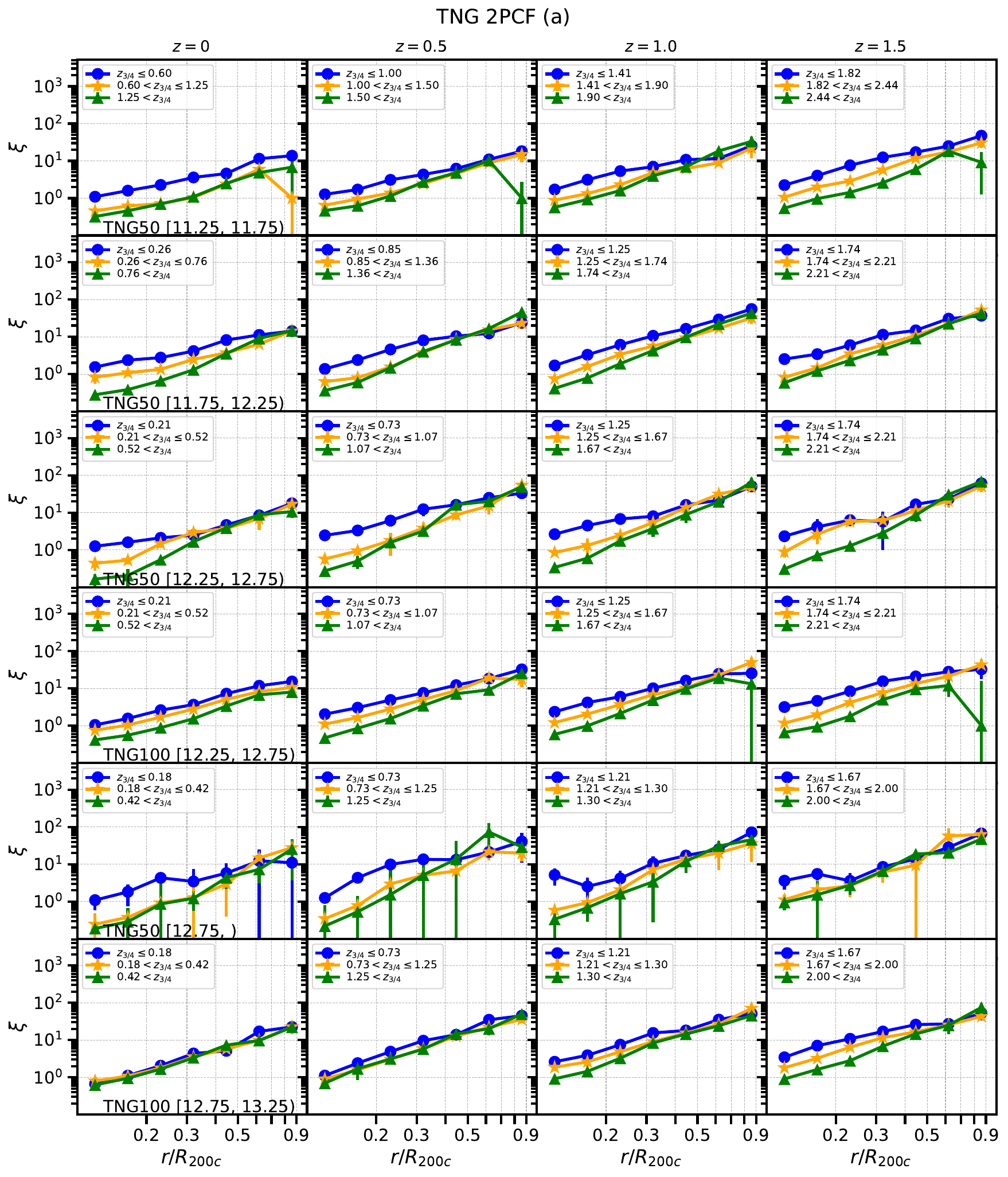}
\caption{This Figure, together with Figure~\ref{fig:TNGcorr_b} below, shows the median 2PCF signals for halo stars in galaxies with different halo mass, redshift and formation redshift from TNG50, TNG100 or TNG300. For panels in the same row, they refer to galaxies with the same host halo mass, and the number in brackets of the left panel in each row corresponds to the log halo mass range. The leftmost panels also indicate the source of the results, specifying whether they come from the TNG50, TNG100 or TNG300 simulation. For panels in the same column, they are 2PCF signals at the same redshift, as indicated by the text on top of each column. In each panel, galaxies are divided into three different ranges according to their formation redshifts ($z_{3/4}$, see the legend), and their median 2PCFs are plotted in different colors and symbols. Here $z_{3/4}$ is defined as the redshift at which one galaxy has accreted three-fourths of its total ex-situ stellar mass at the corresponding redshift. $r$ in the $x$-axis is the galactocentric distance. When calculating the 2PCFs, we choose a fixed pair separation of $\Delta x=2.5-5$~kpc. The error bars are based on the 1-$\sigma$ scatters of 100 bootstrapped sub samples.}
\label{fig:TNGcorr}
\end{figure*}

\begin{figure*}[ht]
\centering
\includegraphics[width=0.86\linewidth]{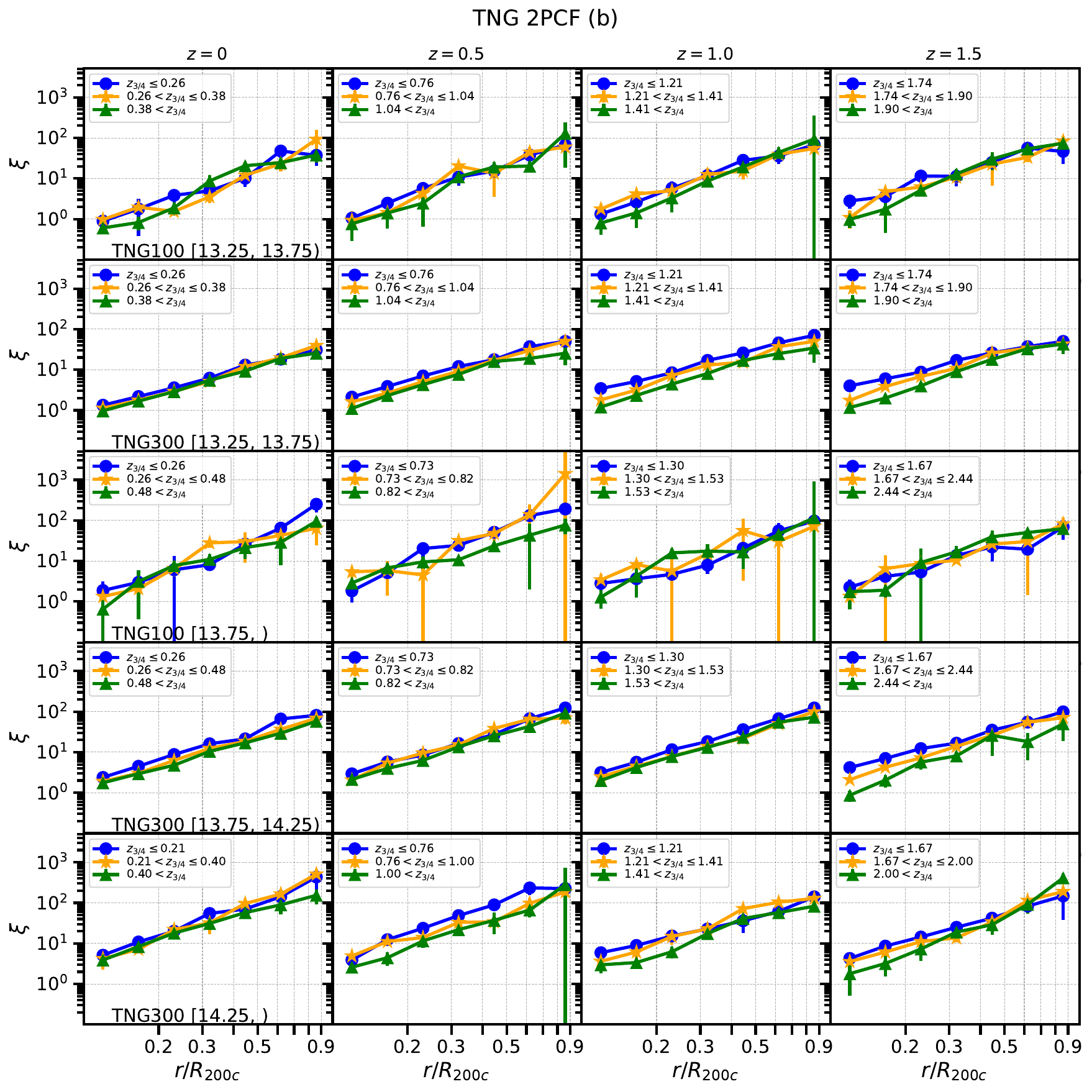}
\caption{A continuation of Figure~\ref{fig:TNGcorr} above. }
\label{fig:TNGcorr_b}
\end{figure*}

Figures~\ref{fig:TNGcorr} and \ref{fig:TNGcorr_b} show the 2PCF signals for halo stars in galaxies from TNG50, TNG100 and TNG300. In the same row, galaxies are in the same bin of host halo virial mass\footnote{For reference, our MW most closely corresponds to the second row ($11.75<\log_{10}M_{200c}/\msun<12.25$), though current measurements of the $M_{200c}$ value of our MW Galaxy can still have large uncertainties. The readers can check \cite{2020SCPMA..6309801W} for a review and a few more recent measurements \citep[e.g.][]{2021MNRAS.501.5964D,2021A&A...654A..25J,2022ApJ...925....1S,2022MNRAS.516..731B,2022MNRAS.511.2610C,2023ApJ...946...73Z,2023ApJ...942...12W,2024MNRAS.528..693O,2024arXiv240811414L}. Different measurements can span a factor of four difference from $M_{200c,\mathrm{MW}}\sim0.5\times10^{12}\msun$ to $\sim2\times10^{12}\msun$. The median value of different measurements is slightly greater than $1\times10^{12}\msun$.}, $M_{200c}$, as indicated by the text on the left of each row (see Table~\ref{tbl:massbins}). In the same column, galaxies are at the same redshift, as indicated by the text on top of each column. We use TNG50, TNG100 and TNG300 to continuously cover the lowest to the highest mass bins, while different TNG simulations can have repeated mass bins, that can be used to check the consistency across different resolutions.

For a given panel of Figure~\ref{fig:TNGcorr} or Figure~\ref{fig:TNGcorr_b}, we divide galaxies into three different bins according to their formation redshifts\footnote{So far, there is no robust constraint on the value of $z_{3/4}$ for our MW, but a few studies have tried machine learning approaches trained using numerical simulations to predict the assembly history related quantities of galaxies from a combination of different observables \citep[e.g.][]{2022MNRAS.515.3938S,2023MNRAS.519.2199E,2023MNRAS.523.5408A}.}, $z_{3/4}$. Here $z_{3/4}$ is defined as the redshift when the galaxies have accreted $3/4$ of their total ex-situ stellar mass at the corresponding snapshot. The three curves with different colors in each panel then show the median 2PCF signals for galaxies in each bin of $z_{3/4}$, with the ranges of $z_{3/4}$ indicated by the legend. Following Paper I, we divide the bins to ensure that there are approximately similar number of galaxies in each.

In particular, we note that the $x$-axis quantity, $r$, of each panel refers to the galactocentric radius or distance, instead of the pair separation adopted in calculating the 2PCFs. And in our calculation, we fix the pair separation to $\Delta x=2.5-5$~kpc following the conclusion of Paper I. We normalize $r$ by the virial radius, $R_{200c}$, of the host dark matter halos. At a given $r$, the correlation signal tells the excess probability of finding pairs of halo stars over the given scales, compared with random distributions. With the increase in $r$, the clustering strength reflected by the 2PCF signals also increases, which is true in all different panels, indicating the outer halo stars with greater galactocentric distances on average maintain stronger spatial clusterings, which is consistent with predictions by other numerical simulations \citep[e.g.][]{2011MNRAS.417.2206C, 2019MNRAS.484.2556L}.

For many panels in Figures~\ref{fig:TNGcorr} and \ref{fig:TNGcorr_b}, we can see a monotonic trend that the clustering strength revealed by the 2PCF signals decreases with the increase in $z_{3/4}$. For galaxies with the most recent or early $z_{3/4}$ values, that means they are assembled the latest or the earliest, they have on average the strongest or the weakest clustering in their halo stars. The blue dots connected by lines have the highest amplitude in most panels, while the green triangles show the lowest amplitudes in many cases, with the orange stars in between. This is consistent with Paper I, in terms that halo stars preserve their initial clusterings better if they are accreted later, though in Paper I we have only looked at MW-mass systems at redshift of $z=0$. 

On top of the general monotonic trend, we can clearly see the mass and redshift dependence of how strongly the 2PCF signals correlate with $z_{3/4}$. With the increase in $M_{200c}$ (going from top to bottom), the differences between the blue dots, orange stars and green triangles get much smaller. We can see that the three curves are better separated in the top four rows, whereas they become more and more overlapped with each other going from top to bottom panels in the same column. On the other hand, with the increase in redshifts (going from left to right), it seems the differences between different color curves slightly increase. This can be seen in the bottom row of Figure~\ref{fig:TNGcorr}, and the second and two bottom rows of Figure~\ref{fig:TNGcorr_b}. In the first row of Figure~\ref{fig:TNGcorr}, curves in the $z=1.5$ panel are also better separated than the other three lower redshift panels.

Comparing the same $M_{200c}$ bin in different resolutions of TNG simulations (the $12.25<\log_{10}M_{200c}/\msun<12.75$ bin for TNG50 and TNG100, and the $13.25<\log_{10}M_{200c}/\msun<13.75$ bin for TNG100 and TNG300), we can see they in general show the same trend, in terms of both the overall amplitudes and changes in the 2PCF signals with $z_{3/4}$, $M_{200c}$ and redshifts. The results are, however, not identical. For example, the orange stars and green triangles tend to be lower in the $12.25<\log_{10}M_{200c}/\msun<12.75$ panels of TNG50 than the corresponding TNG100 panel at $r<0.3R_{200c}$, and for the same $M_{200c}$ bin, the three curves are slightly better separated at $r>0.3R_{200c}$ in TNG100 than TNG50. The lower amplitudes at $r<0.3R_{200c}$ for the orange stars and green triangles of TNG50 than TNG100 might be related to the resolution. There are more smaller satellites in the higher resolution TNG50 run, which are disrupted faster in host halos with the same mass. However, such small differences may also be due to the cosmic variance in different TNG simulations. In particular, TNG50 is not a subvolume from TNG100. Nevertheless, we note that the general trend is consistent. The amounts of separations between different symbols/curves are similar. In the end, we note that for the most massive $M_{200c}$ bins of TNG50 and TNG100, the bin widths are chosen to be wide, but the measured 2PCF signals are quite noisy, given the small number of massive galaxies in the smaller box. The massive end is better probed with TNG300, with its largest box size.

\begin{figure*}
\centering
\includegraphics[width=1\linewidth]{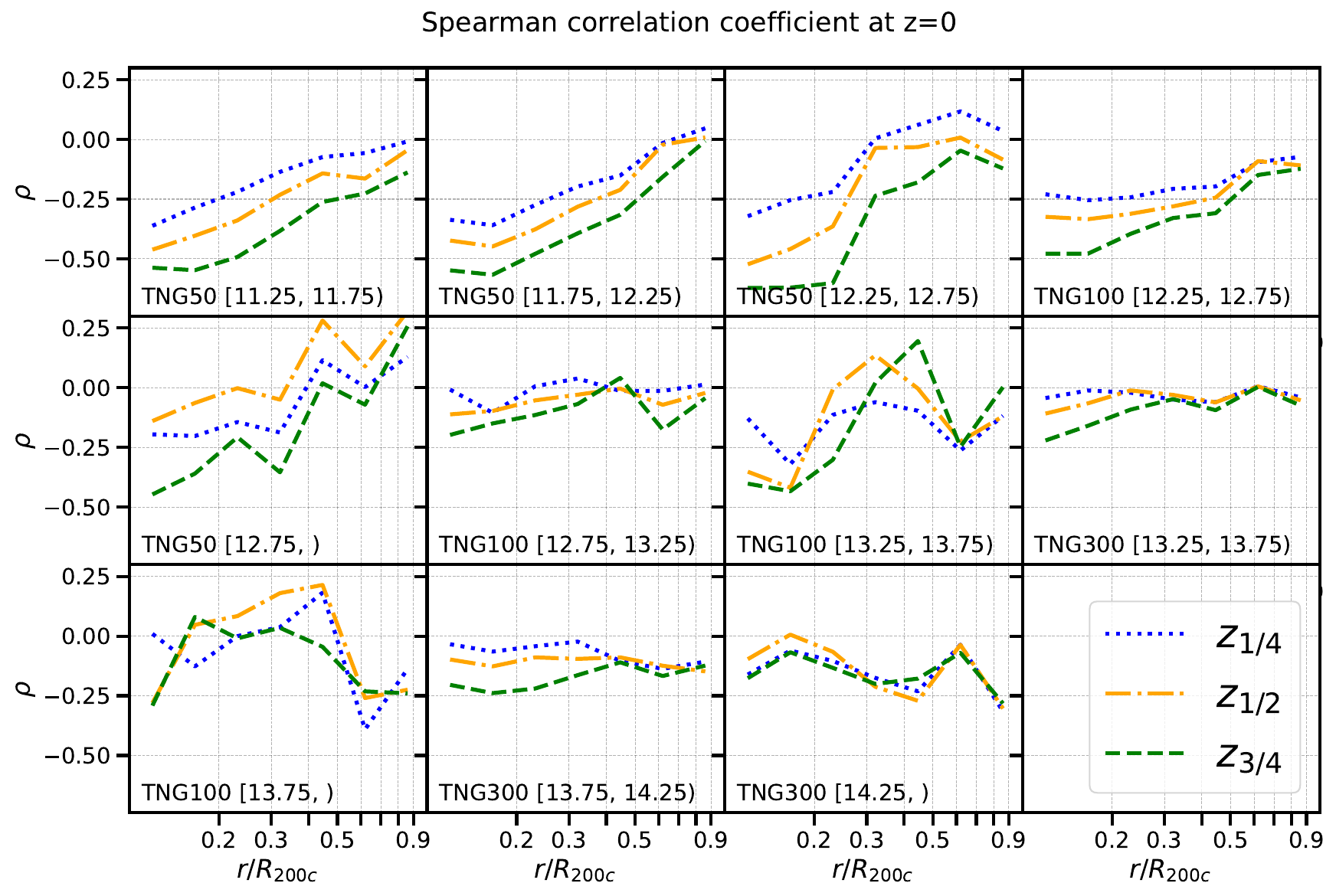}
\caption{The Spearman correlation coefficients ($\rho$) between the 2PCF signals and the galaxy formation redshifts defined in different ways and at different galactocentric distances. Following Figure~\ref{fig:TNGcorr}, we adopt TNG50, TNG100 and TNG300 to continuously cover the full mass range, with the name of the simulation and the mass range of $\log_{10}M_{200c}/\msun$ indicated by texts of each panel. We only show the results at redshift $z=0$. In each panel, dotted blue, dash-dotted orange and dashed green curves refer to the Spearman correlation coefficients with $z_{1/4}$, $z_{1/2}$ and $z_{3/4}$. Here $z_{1/4}$, $z_{1/2}$ and $z_{3/4}$ are defined as the redshifts at which the galaxies have accreted one-fourth, half and three-fourths of their total ex-situ stellar mass at redshift $z=0$. Negative correlation coefficients mean the clustering strength revealed by the 2PCF signals and the galaxy formation redshifts are anti-correlated, i.e., the clustering decreases with the increase in the formation redshifts.}
\label{fig:TNGrho}
\end{figure*}

Figure~\ref{fig:TNGrho} further shows the Spearman correlation coefficients between the 2PCF signals at different galactocentric distances and the formation redshifts of galaxies. We only show the $z=0$ result to avoid redundancy. Following Paper I, we adopt three different galaxy formation redshifts, with $z_{1/4}$, $z_{1/2}$ and $z_{3/4}$ defined as the redshifts at which the galaxies have accreted one-fourth, half and three-fourths of their total ex-situ stellar mass at the corresponding snapshot, respectively. Consistent with Paper I, we see the most negative correlations for $z_{3/4}$ (dashed green curve), and the weakest correlations for $z_{1/4}$ (dotted blue curve). This is because $z_{3/4}$ quantifies those halo stars that are accreted later, and they are expected to preserve their clusterings better. This is also the reason why we choose to use $z_{3/4}$ in Figures~\ref{fig:TNGcorr} and \ref{fig:TNGcorr_b} above.

According to Figure~\ref{fig:TNGrho}, we can also see the trend over different mass bins. In general,  the correlation coefficients are more negative in the first row. The correlation coefficients are significantly closer to zero in the other more massive panels. This is consistent with Figures~\ref{fig:TNGcorr} and \ref{fig:TNGcorr_b}.

\section{Discussions}
\label{sec:disc}

\subsection{Phase mixing versus new accretion}
\label{sec:mixandacc}

In the previous section, we have presented our main results of how does the correlation between the 2PCF signals and the galaxy formation redshifts depend on the host halo mass and redshift. We see some weak dependence on redshift, in terms that at higher redshifts the correlation gets slightly stronger. This might reflect the weakened clustering in accreted particles due to phase mixing as time passes, resulting in weaker correlations. Moreover, we also detected dependence on the host halo mass, in terms that with the increase in the virial mass of host dark matter halos, the correlation between the 2PCF signals and the galaxy formation redshifts weakens. In this subsection, we try to further explore the connections between the 2PCF signals and the mass assembly histories, before providing more detailed discussions later in Section~\ref{sec:massdependence_reason} about the explanations to the halo mass dependence shown from Figures~\ref{fig:TNGcorr} to \ref{fig:TNGrho}.

\begin{figure*}
\centering
\includegraphics[width=1\linewidth]{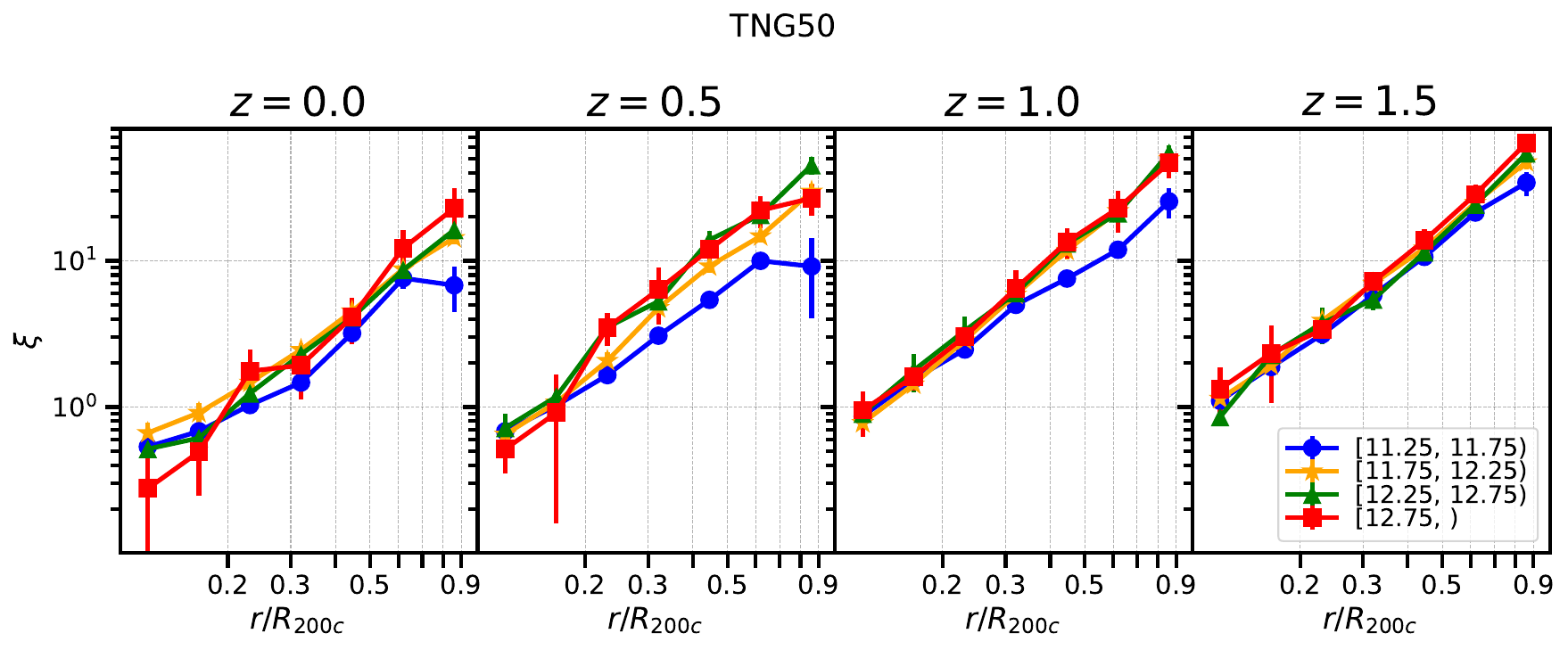}
\caption{The 2PCF signals as a function of galactocentric distances, $r$, scaled by $R_{200c}$. This figure is based on TNG50. We plot in each panel 2PCF measurements at the same redshift/snapshot. Different symbols connected by lines with varying colors in the same panel refer to the median 2PCFs for galaxies in four mass bins of $M_{200c}$ (see the legend). Galaxies are no longer divided into bins according to their formation redshifts.}

\label{fig:massbin_compare}
\end{figure*}

We investigate how the absolute amplitudes of 2PCF signals depend on host halo mass and redshifts. Symbols and lines in panels of Figure~\ref{fig:massbin_compare} are median 2PCFs for galaxies at the same redshift, but in different host halo mass bins, as indicated by the legend. Note unlike Figure~\ref{fig:TNGcorr}, we no longer divide galaxies into bins according to their formation redshifts. In a given panel, the difference between different symbols connected by lines with varying colors is not large. The red squares, green triangles and orange stars connected by lines with corresponding colors are very close to each other, but we can see some trend that the 2PCF signals in more massive bins are slightly higher in amplitude. Note, however, part of the red squares and green triangles drop below orange stars in the $z=0$ and $z=0.5$ panels. The blue dots of the least massive bin more prominently drop below the others.

\begin{figure}
\centering
\includegraphics[width=1\linewidth]{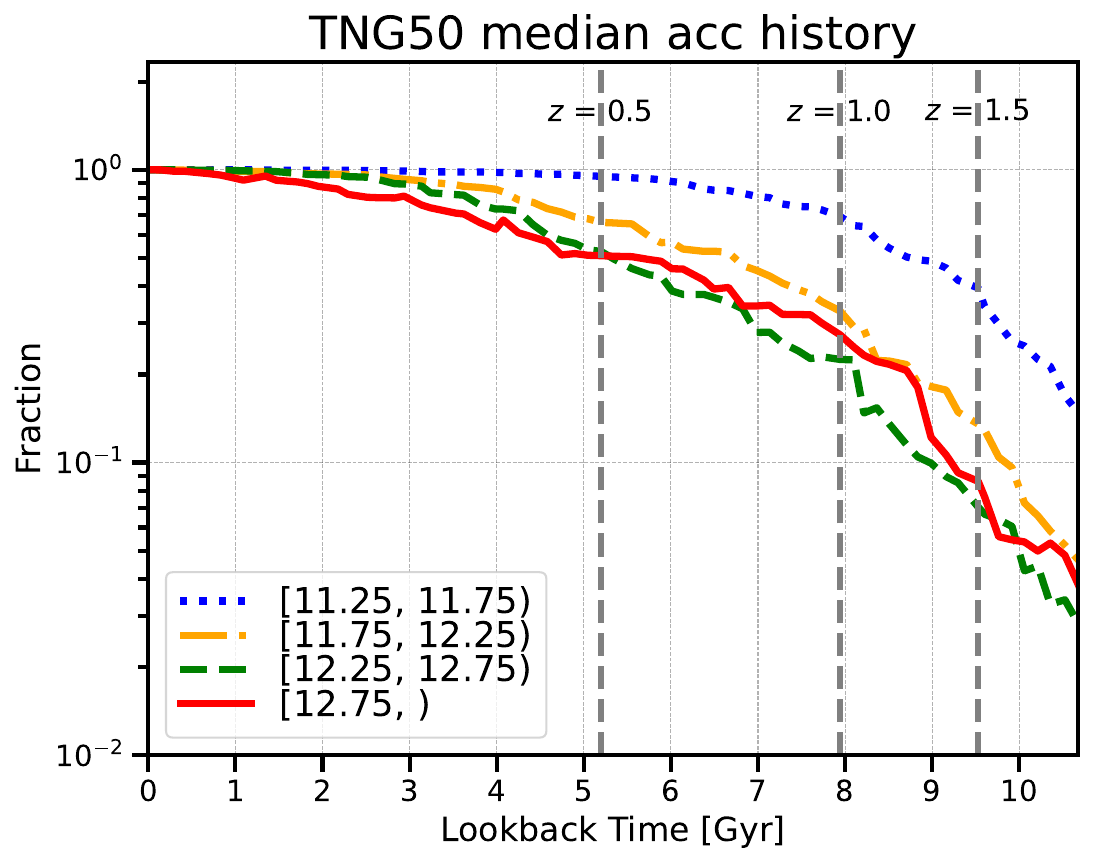}
\caption{The ex-situ stellar mass accretion histories of galaxy systems in four different halo mass ($M_{200c}$) bins, as indicated by the legend. The curves are medians taken over the assembly histories of galaxies in the same mass bin, and are normalized to be unity at redshift $z=0$. The gray dashed vertical lines mark the positions of redshifts $z=0.5$, $1.0$ and $1.5$. The plot is based on TNG50.}
\label{fig:mean_acc}
\end{figure}

It is well recognized that more massive halos/galaxy systems assemble later in the universe \citep[e.g.][]{2019MNRAS.487.5416T, 2019MNRAS.488.3143B, 2023ApJ...953...37G}, and as we have checked, more massive galaxy systems in TNG50 are still accreting a more significant fraction of their mass at today below redshift $z=1$ than less massive systems\footnote{This is in fact also reflected in Figures~\ref{fig:TNGcorr} and \ref{fig:TNGcorr_b}, that when we go to higher mass bins, the values of $z_{3/4}$ we adopted to divide galaxies into different bins of formation redshifts become smaller and smaller. }. The median ex-situ stellar mass accretion histories for galaxies in the four $M_{200c}$ bins are shown in Figure~\ref{fig:mean_acc}. For the lowest mass bin, the growth flattens at $z<1$. For the other three more massive bins, they are still more actively growing at $z<1$. 

The newly accreted stars would increase the overall spatial clusterings in halo stars. Hence increasing the 2PCF signal amplitudes for more massive systems. In fact, as we are calculating and plotting the 2PCF signals for all halo stars in Figure~\ref{fig:massbin_compare}, the amplitudes of the signals are modulated by two factors: 1) the clustering strength gets weakened with time, due to the reduction in spatial clustering of early accreted stars (phase mixing); 2) The newly added material would further increase the clustering. The overall clustering is a net outcome of both 1) and 2). More massive galaxy systems accrete larger fractions of mass at lower redshifts, and we think this is the reason why curves for the three more massive systems have higher amplitudes than the least massive bin in Figure~\ref{fig:massbin_compare}. The difference between the three curves in more massive halos is very small, because they have more similar assembly histories (Figure~\ref{fig:mean_acc}).

\begin{figure*}
\centering
\includegraphics[width=1\linewidth]{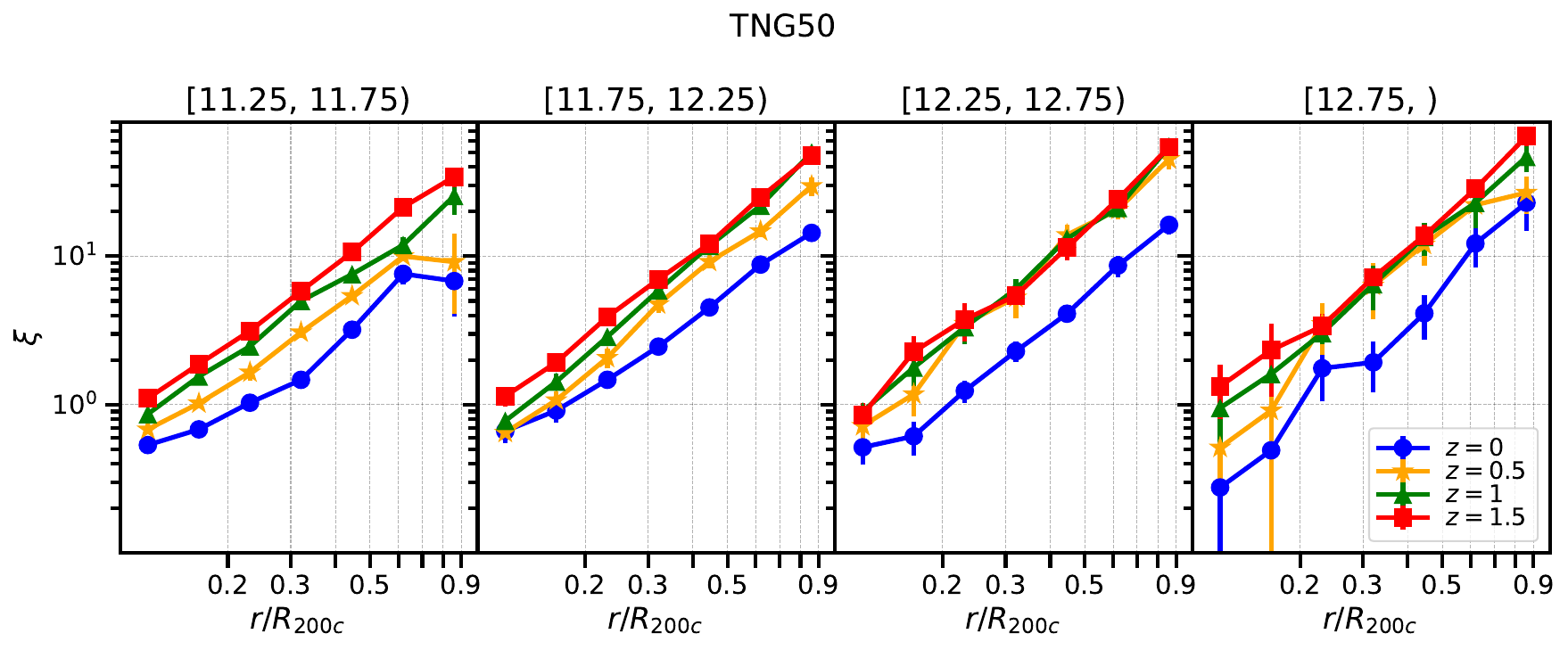}
\caption{Similar to Figure~\ref{fig:massbin_compare}, but each panel shows the median 2PCFs in the same mass bin of $M_{200c}$, while different color curves refer to the measurements at different redshifts (see the legend).
}
\label{fig:zbin_compare}
\end{figure*}

Figure~\ref{fig:zbin_compare} is similar to Figure~\ref{fig:massbin_compare}, but we further show the 2PCFs in the same mass bin of $M_{200c}$ in each panel, and use different symbols connected by lines with varying colors to compare the measurements at four redshifts. With the decrease in redshift, the clustering amplitudes get reduced, indicating the spatial clustering of halo stars gets weakened due to phase mixing as time passes. However, the trend is less clear in more massive bins. The orange stars, green triangles and red squares almost overlap with each other in the two more massive panels, and are also closer to each other in the second least massive bin, whereas they become much better separated in the least massive bin (the left panel).

The trends in Figure~\ref{fig:zbin_compare} are the combined outcome of the two factors that we mentioned above. The slopes for curves in Figure~\ref{fig:mean_acc} reflect the relative increase in ex-situ stellar mass with respect to time. For the three more massive bins, they are still actively growing at $z>0.5$, and the growing rate becomes smaller at $z<0.5$. At $z>0.5$, the more active growth counteracts with the effect of phase mixing, resulting in similar amplitudes in the 2PCF signals for the $z=1.5$, $z=1$ and $z=0.5$ curves in the two more massive bins of Figure~\ref{fig:zbin_compare}. We note that the red solid, green dashed and orange dot dashed curves in Figure~\ref{fig:mean_acc} are close to each other, with the orange dot dashed curve slightly more flattened than the other two, while they all show significantly steeper slopes than that of the blue dotted curve. In the same time, we also see the orange stars, green triangles and red squares are close to each other in the three more massive panels of Figure~\ref{fig:zbin_compare}, with those in the $11.75<\log_{10}M_{200c}/\msun<12.25$ panel (corresponding to the orange dot dashed curve in Figure~\ref{fig:mean_acc}) show some better separation. These trends are all self-consistent. For the least massive bin, its growth in halo stars is significantly less than the other three more massive bins, and thus the least massive panel in Figure~\ref{fig:zbin_compare} is dominated by factor 1) above, i.e., phase mixing, and thus the red squares, green triangles, oranges stars and blue dots drop steadily and are better separated. In the end, we note that the blue dots in the two most massive panels of Figure~\ref{fig:zbin_compare} drop significantly compared with the other three curves. This is associated with the flattening (slower relative mass growth) of the corresponding curves in Figure~\ref{fig:mean_acc} at $z<0.5$. The decrease in the growth rate of ex-situ stars over time is generally expected as halo growth is known to transit from a fast growth phase to a slow growth phase over time~\citep{Zhao,2023ApJ...953...37G}, with massive halos lagging behind low mass halos in this process. In addition, the mass loss rates of subhalos also respond to the two-phase halo mass evolution, with earlier accreted subhalos suffering from more rapid mass loss~\citep{Feihong}.

\begin{figure}
\centering
\includegraphics[width=1\linewidth]{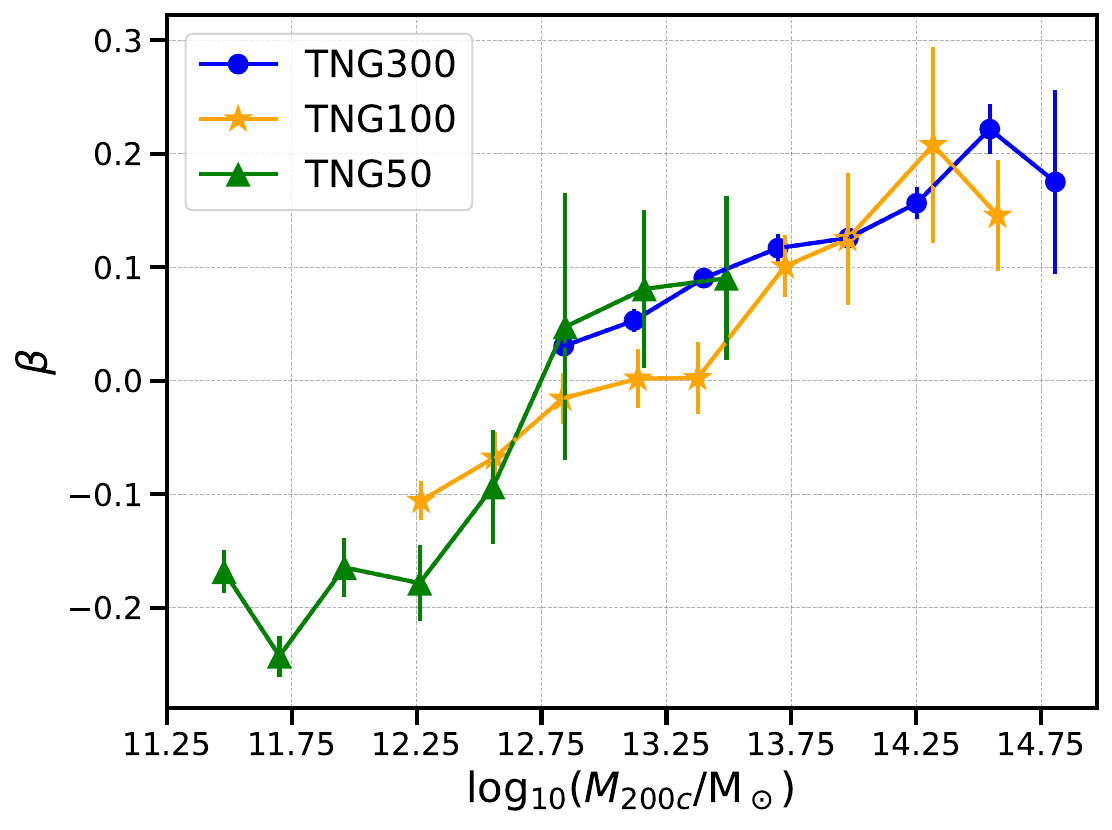}
\caption{The velocity anisotropy ($\beta$) of subhalos, different symbols and colors indicate different simulations. Here $\beta$ is calculated using subhalos with galactocentric distances of $0.1R_{200c}<r<R_{200c}$. The number of subhalos within a single halo in this range should exceed 10 as well as the number of dark matter particles within a single subhalo should exceed 20 in order to exclude the effect of small mass dark matter structures.}
\label{fig:beta_mass}
\end{figure}

\subsection{Why does the correlation between 2PCF signals and formation redshifts depend on halo mass?}
\label{sec:massdependence_reason}

According to Figures~\ref{fig:massbin_compare} to ~\ref{fig:zbin_compare} in Section~\ref{sec:mixandacc} above, it is clearly demonstrated that the clustering or 2PCF signal strength in halo stars is modulated by the two factors mentioned: phase mixing to decrease the clustering and newly accreted stars to increase the clustering. 
Now we would like to discuss possible explanations to the question: why are the correlations between halo star 2PCF signals and the formation redshifts weaker in massive halos? We discuss possible reasons below:

\subsubsection{Tidal effects}

It is clear that massive galaxy systems are still more actively accreting than less massive systems. Then is it possible that halo star clusterings are also disrupted more efficiently in more massive galaxy systems?
Maybe the clusterings in accreted halo stars are destroyed more efficiently, hence losing the correlation about the infall redshifts faster, whereas halo stars in less massive galaxy systems can maintain their spatial clusterings for longer time.

What might be responsible for causing stronger destruction in halo star clustering in more massive systems? One may think about tidal effects, which is stronger in more massive dark matter halos. More massive dark matter halos have stronger tidal effects on progenitor satellite galaxies with the same mass and orbit. However, the satellites of more massive hosts are also more massive, so tidal effect is in principle scale invariant, which cannot directly explain the mass dependent trends that we see in previous figures.

\subsubsection{Infalling orbits of satellites}

Another possible explanation might be related to the infalling orbits of satellite galaxies. The infalling orbits of satellite galaxies may affect the rate of disruption and phase mixing. Satellites with more radial orbits upon accretion are expected to be more strongly and quickly disrupted, as they can approach the halo center faster, sometimes even directly cross the central galaxy, losing most of their bound dark matter and stars. We also expect the stripped halo stars from satellites with more radial orbits to be dynamically hot and phase mix more quickly. According to our results, one might anticipate that satellites or subhalos in more massive systems fall in on more radial orbits.

The explanation above seems to be supported by Figure~\ref{fig:beta_mass}, where we show the velocity anisotropy parameter of subhalos from TNG50, TNG100 and TNG300. We again adopt different resolutions of TNG simulations to cover a wide mass range. The velocity anisotropy parameter is defined in Section~\ref{sec:beta}, and we calculate the velocity anisotropy using subhalos with galactocentric radii between 0.1$R_{200c}$ and $R_{200c}$. We avoid the very central region within $0.1 R_{200c}$, which is sensitive to baryonic physics, and subhalos in inner regions are severely stripped. Larger values of $\beta$ mean more radial orbits, and clearly, we can see that $\beta$ becomes more radial with the increase in $M_{200c}$, meaning satellites/subhalos accreted by more massive dark matter halos have more radial orbits\footnote{Our measurement is consistent with \cite{2020ApJ...905..177L} and \cite{2024arXiv240714827H}. The more radial orbits of satellites in more massive halos are consistent with the fact that massive halos are still more actively accreting. It has been shown that massive halos are surrounded by an active infall region with prominent radial motions \citep[e.g.][]{FH21,Gao23,Diemer22}.}.

However, the explanation that the more radial orbits of satellites in more massive systems cause stronger disruption and phase mixing cannot qualitatively explain Figure~\ref{fig:massbin_compare}. Galaxy systems in the three more massive bins have similar ex-situ stellar mass assembly histories as shown by Figure~\ref{fig:mean_acc}, meaning that they have similar growth rate of newly added halo stars to increase the 2PCF signals. Nevertheless, the significant increase in $\beta$ with host halo mass indicates that the 2PCF signals would have been weakened more quickly in massive bins, and thus we would expect to see the red squares having lower amplitudes than the green triangles, and the green triangles having lower amplitudes than the orange stars. This is only seen for a few data points in the $z=0$ panel, that the red squares are slightly below the orange stars and green triangles at $r<0.2R_{200c}$ and $r\sim0.3R_{200c}$, and the green triangles are slightly below the orange stars at $r<\sim0.6R_{200c}$. The differences are, however, very small. 

\subsubsection{Mass dependent accretion rate}

In the end, we propose a more plausible explanation, that is related to the mass dependent accretion rate. The amount of newly accreted halo stars, which increases the clustering or 2PCF signal strength, is determined by the more recent mass accretion rate. On the other hand, the degree of phase mixing, which decreases the clustering in halo stars, is determined by the overall accretion history and formation redshifts over a much longer time scale. As we have already shown in Figure~\ref{fig:mean_acc} above, for low mass halos, they are no longer actively accreting, and their halo stars were mostly accreted earlier, which then show steady decrease in clustering strength as time passes and due to phase mixing (the left panel of Figure~\ref{fig:zbin_compare}). Hence, low mass halos show stronger correlation between the 2PCF signal of their halo stars and formation redshifts. On the other hand, for massive halos, they are more actively accreting. Stars accreted earlier by more massive halos are undergoing phase mixing, but the large amount of newly accreted stars does not yet have time to mix. This results in a masking of the correlation between 2PCF signals and formation redshifts, as formation redshift characterizes mass accretion over a much longer timescale.

\section{Conclusions}
\label{sec:concl}

In this paper, we adopt the two point correlation function (2PCF) as a tool to quantify the spatial clustering of accreted halo star particles in the IllustrisTNG suites of simulations. In a previous study \citep[][Paper I]{2024ApJ...961..223Z}, we investigated the correlation between the strength of 2PCF signals for halo stars and the galaxy formation redshifts, for MW-mass galaxy systems from TNG50. We reported clear correlations, but the scatter in the correlations is too large to enable precise inference of the galaxy formation redshifts with 2PCF signals along. In this study, we move on by extending the studies in Paper I to a wider range in galaxy mass and redshifts, by jointly using TNG50, TNG100 and TNG300 simulations. 

Despite the large scatters reported in Paper I, we have detected that the correlation between the 2PCF signal strength and the galaxy formation redshifts depends on both halo mass and redshifts, for galaxy systems covering a wide host halo mass range ($11.25<\log_{10}M_{200c}/\msun<15$) and redshift range ($0<z<1.5$). In general, the correlation gets slightly stronger at higher redshifts, and weakens in more massive galaxy systems, indicating it is more difficult to infer the mass assembly histories for more massive galaxy systems using 2PCF statistics of halo stars.

We demonstrate that the spatial clustering of halo stars, as reflected in our 2PCF signals, is affected by two factors: 1) As time passes, the spatial clustering in accreted halo stars weakens due to phase mixing; 2) newly accreted stars at more recent time increase the spatial clustering. For more massive galaxy systems, they assemble later, so they are still more actively accreting new halo stars at today. 

The late assembly of massive systems may help to explain the weaker correlations between the 2PCF signals and the galaxy formation redshifts in massive halos, as their 2PCFs are affected more by recently accreted stars while formation redshift characterizes mass accretion on a much longer timescale. The orbits of satellite galaxies in more massive halos also maintain a larger radial anisotropy, reflecting the more active accretion state of their hosts.

\acknowledgments
This work is supported by NSFC (12273021, 12022307), the National Key R\&D Program of
China (2023YFA1605600, 2023YFA1605601), the China Manned Space (CSST) 
Project with No. CMS-CSST-2021-A02 and No. CMS-CSST-2021-A03, the National Key Basic Research and Development Program of China (No.~2018YFA0404504), 111 project (No.~B20019) and Shanghai Natural Science Foundation (No. 19ZR1466800). We thank the sponsorship from Yangyang Development Fund. Z-L acknowledges the funding from the European Unions Horizon 2020 research and innovation programme under the Marie Skodowska-Curie grant 101109759 ("CuspCore"). J-H acknowledges the National Key R\&D Program of China (2023YFA1607800, 2023YFA1607801). The computation of this work is carried out on the \textsc{Gravity} supercomputer at the Department of Astronomy, Shanghai Jiao Tong University. We are very grateful for useful discussions with Dandan Xu. We thank the anonymous referee for his/her time spent on reading and commenting this paper. 

%\vspace{5mm}
%\facilities{HSS-SSP}
%\software{}

%\clearpage

\bibliography{master}{}
\bibliographystyle{aasjournal}

\clearpage



\end{document}